\documentclass[aps,prb,twocolumn,showpacs]{revtex4}
\usepackage{amssymb}

\usepackage{graphicx}

\begin{document}

\title{Distinct behaviors of suppression to superconductivity in $LaRu_3Si_2$ induced by Fe and Co dopants}

\author{Sheng Li,  Jian Tao, Xiangang Wan, Xiaxin Ding, Huan Yang and Hai-Hu Wen}\email{hhwen@nju.edu.cn}

\affiliation{Center for Superconducting Physics and Materials,
National Laboratory of Solid State Microstructures and Department
of Physics, Nanjing University, Nanjing 210093, China}

\begin{abstract}
In the superconductor LaRu$_3$Si$_2$ with the Kagome lattice of
Ru, we have successfully doped the Ru with Fe and Co atoms.
Contrasting behaviors of suppression to superconductivity is
discovered between the Fe and the Co dopants: Fe-impurities can
suppress the superconductivity completely at a doping level of
only 3$\%$, while the superconductivity is suppressed slowly with
the Co dopants. A systematic magnetization measurements indicate
that the doped Fe impurities lead to spin-polarized electrons
yielding magnetic moments with the magnitude of 1.6 $\mu_B$\ per Fe,
while the electrons given by the Co dopants have the same density of states for spin-up and spin-down leading to much weaker magnetic moments. It is the
strong local magnetic moments given by the Fe-dopants that
suppress the superconductivity. The band structure calculation
further supports this conclusion.
\end{abstract}

 \maketitle

\section{Introduction}

Superconductivity in the systems $RT_3$Si$_2$ or $RT_3$B$_2$ ($R$
stands for the rare earth elements, like La, Ce, Y, etc., $T$ for
the transition metals, like Ru, Co and Ni, etc.) is very
interesting because it concerns the conduction of the d-band
electrons of the 3d- or 4d- transition metals. By having different
combinations of chemical compositions, one can tune the system
from a superconducting (SC) ground state to a magnetic one, and
sometimes have both phases coexisting in one single
sample.\cite{Escorne,KuHQ} The LaRu$_3$Si$_2$ has a SC transition
temperature at about 7.8 K.\cite{Barz,Godart} Since the
superconductivity is at the vicinity of the magnetic order, some
unconventional pairing mechanisms, such as the charge
fluctuation\cite{Rauchschwalbe}, or anti-ferromagnetic spin
fluctuations\cite{Scalapino} mediated pairings are possible.
Recently, we find that both the superconducting state and the
normal state exhibits some anomalous properties, suggesting that
the electronic correlation plays important roles in the occurrence
of superconductivity.\cite{LiShengPRB} Another reason for doing
research on this system is that it may have some odd pairing
symmetries, such as $d-wave$, $s+d$, $p_x+ip_y$ or
$d_{x^2-y^2}+id_{xy}$,\cite{WangZDPRL,LiJX} because the electric
conduction is dominated by the 4d band of $Ru$ atoms which
construct a Kagome lattice (a mixture of the honeycomb and
triangular lattice, as shown in Fig.1). Furthermore, the electric
conduction in this system is strongly favored by the Ru-chains
along the z-axis, as evidenced by our band structure calculations,
this may induce quite strong superconducting
fluctuations.\cite{LiShengPRB}

In a superconductor, the impurity induced pair breaking depends
strongly on the structure of the pairing gap and the feature of
the impurities, such as magnetic or non-magnetic. Therefore it is
very important to measure the impurity induced scattering effect
in the superconducting state of $LaRu_3Si_2$. According to the
Anderson's theorem,\cite{AndersonTheorem,Balatsky} in a
conventional s-wave superconductor, nonmagnetic impurities will
not lead to apparent pair-breaking effect. This theoretical
expectation has been well illustrated in the conventional
superconductors.\cite{Yazdani}However, a magnetic impurity, due to
the effect of breaking the time reversal symmetry, can break
Cooper pairs easily. In sharp contrast, in a d-wave
superconductor, nonmagnetic impurities can significantly alter the
pairing interaction and induce a high density of states (DOS) due
to the sign change of the gap on a Fermi surface. This was indeed
observed in cuprate superconductors where Zn-doping induces
T$_c$-suppression as strong as other magnetic disorders, such as
Mn and Ni.\cite{PanSH,CuprateZnMn} In LaRu$_3$Si$_2$, preliminary
experiment indicated that the SC transition temperature drops only
1.4 K with the substitution of 16 $\%$ La by Tm (supposed to
possess a magnetic moment of about 8$\mu_B$), suggesting that the
superconductivity is robust against the local paramagnetic
moment\cite{Escorne}. This kind of doping is induced at the sites
of the rare earth elements, which may give very weak pair breaking
effect on the Cooper pairs (3d electrons of $Ru$). Therefore it is
very interesting to investigate what will happen if we dope
impurity atoms directly to the $Ru$ sites. In this paper, we report
the doping effect on the $Ru$ sites by the Fe and Co dopants. We
find a contrasting suppression effect to the superconductivity
with these two kind of dopants. Possible reasons are given to
explain this effect.

\begin{figure}
\includegraphics[width=9cm]{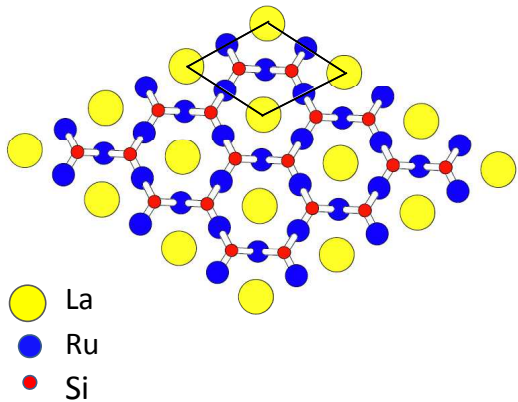}
\caption {(color online) Top view of the atomic structure of LaRu$_3$Si$_2$.
The Ru atoms construct a Kagome lattice (blue middle size circles), while the Si (red small size circles) and La atoms (yellow large size circles)
form a honeycomb and a triangle structure, respectively. The three
different atoms don't overlap each other from a top view. The
prism at the top corner illustrates one unit cell of the
structure. } \label{fig1}
\end{figure}

\section{Experimental methods and characterization}

\begin{figure}
\includegraphics[width=9cm]{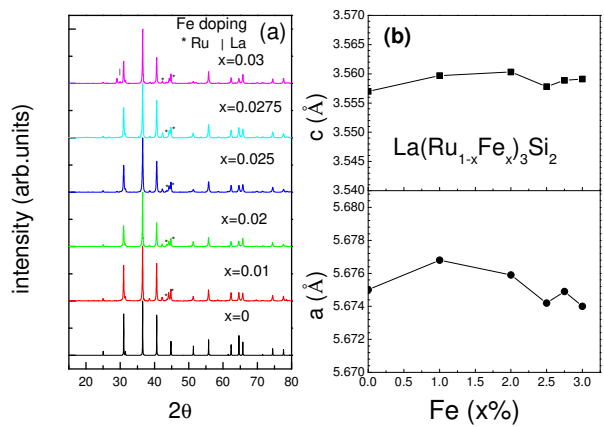}
\caption {(color online) (a) X-ray diffraction patterns of the
sample La(Ru$_{1-x}$Fe$_{x}$)$_{3}$Si$_2$. One can see that the main
phase is the 132 structure, with slight Ru impurity phase. (b) Doping
dependence of the a-axis and c-axis lattice constants. Because Fe
doping is only up to 3$\%$, so there is no distinct change of the
lattice constant.}\label{fig1}
\end{figure}

The samples of La(Ru$_{1-x}$T$_{x}$)$_{3}$Si$_2$ (T=Fe and Co)
were fabricated by the arc melting
method.\cite{Escorne,Barz,Godart,LiShengPRB} The starting
materials: La metal pieces (99.9$\%$, Alfa Aesar), Fe powder
(99.99$\%$), Co powder (99.99$\%$), Ru powder (99.99$\%$) and Si
powder (99.99$\%$) were weighed, mixed well, and pressed into a
pellet in a glove box filled with Ar (water and the oxygen
compositions were below 0.1 PPM). In order to avoid the formation
of the LaRu$_2$Si$_2$ phase, we intentionally add a small amount
of extra Ru powder (about 15$\%$ more) in the starting materials.
Three rounds of welding with the alternative upper and bottom side
of the pellet were taken, in order to achieve the homogeneity. The
X-ray diffraction (XRD) measurement was performed on the Brook
Advanced D8 diffractometer with Cu K$_\alpha$ radiation. The
analysis of XRD data was done with the softwares Powder-X and
Fullprof. The resistivity and magnetization measurements were done
on the Quantum Design physical property measurement system
(PPMS-16T) and SQUID-VSM.

The XRD patterns for Fe- and Co- doped samples are shown in Fig.2
and Fig.3, respectively. One can see that the samples are rather
clean, except for a small amount of Ru impurity. For the Fe-doped
samples, we don't see clear change of the lattice constant $a$ and
$c$. This could be due to two reasons: (1) The maximal doping
level here is 3$\%$ which is already enough to kill the
superconductivity completely; (2) The atoms of Fe and Ru are in
the same column and close to each other in the periodic table, we
would assume that the ions of Fe and Ru have the similar radii.
For the Co-doping, however, there is an obvious decrease of $a$
and $c$ lattice constant with doping, as shown in Fig.3. The
variation of the lattice constants in the Co-doped samples are
well associated with the resistivity data shown below, clearly
suggesting that the Co atoms are also successfully doped into the
LaRu$_3$Si$_2$ system.

\section{Results and Discussion}
\begin{figure}
\includegraphics[width=9cm]{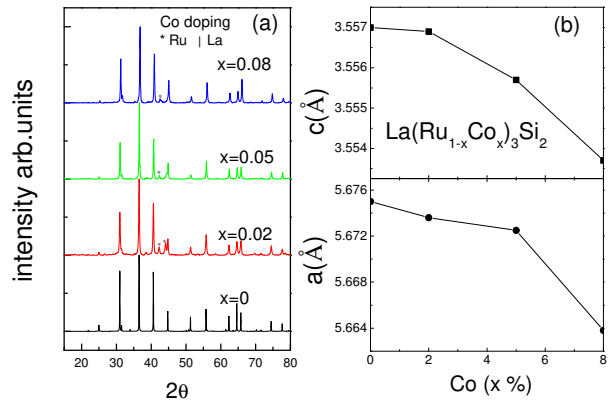}
\caption {(color online) (a) X-ray diffraction patterns of the
sample La(Ru$_{1-x}$Co$_{x}$)$_{3}$Si$_2$, up to the doping level
of 8$\%$ the sample is still quite clean. (b) Doping
dependence of both $a$ and $c$ lattice constants with the increase
of doped Co concentration.} \label{fig3}
\end{figure}

\begin{figure}
\includegraphics[width=9cm]{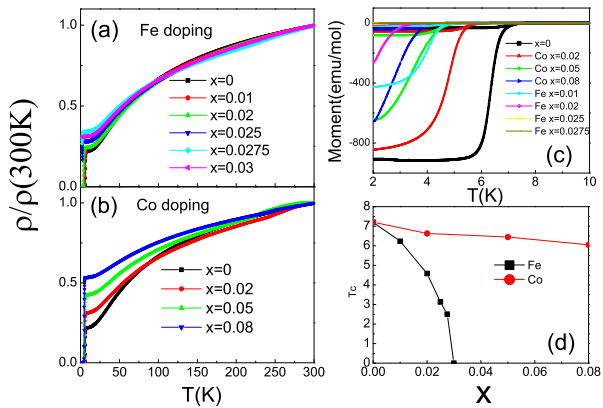}
\caption {(color online) (a) Temperature dependence of normalized
resistivity with Fe doping, there is no superconducting transition
with the doping level at only 3$\%$. Slightly enhancement of the
residual resistivity is observed, indicating an enhanced
scattering. (b) Temperature dependence of the normalized
resistivity with Co doping, the suppression to the superconducting
transition by Co doping is rather weak. (c) Temperature dependence
of DC susceptibility of the La(Ru$_{1-x}$T$_{x}$)$_{3}$Si$_2$
(T=Fe and Co) under H = 50 Oe, measured in the zero-field-cooled
(ZFC) and field-cooled (FC) processes. (d) Doping dependence of
T$_c$ in Fe- and Co-doped samples, the suppression to T$_c$ in Fe
doped samples is drastically fast, but that by Co doping is very
slow.} \label{fig4}
\end{figure}

\subsection{Suppression to superconductivity}
In Fig.4(a) and (b), we present the temperature dependence of the
normalized resistivity of Fe- and Co- doped samples. It can be
seen that the transition temperature was suppressed remarkably
with Fe-doping, and shifted to below 2 K at only a doping level of
$3\%$. However, for the Co-doped ones, there is no significant
change of T$_c$, up to $8\%$ Co-doping. These behaviors are also
revealed by the magnetization of the samples, as shown in
Fig.4(c). For the superconducting samples, the resistivity
increases monotonously with the increase of doping level, both for
the Fe and Co doping. But it is clear that, the enhancement of the
residual resistivity, although weaker in the Fe-doped samples than
that in the Co-doped ones, but the suppression to the
superconductivity is the opposite. In Fig. 4(d) we illustrate the
suppression of T$_c$ with doping concentrations of Fe and Co. This
is easy to be understood in the way that, the suppression to the
superconductivity in the Fe-doped samples is induced by the local
magnetic moments. These magnetic scattering centers are detrimental to the
Cooper pairs, and thus suppress the superconducting transition
temperature significantly. However, in the normal state these
impurities, although possessing strong magnetic moments, act as
the usual scattering centers. In the Co-doped case, the increase
of the residual resistivity is quite strong. For example, the
residual resistivity increases more than 100 $\%$ with the Co
doping level of about 8$\%$. However, the superconducting
transition temperature drops only about 2 K. This sharp contrast
between the behaviors of the Fe and Co-doped samples is unexpected
from a straight forward picture, since both Fe and Co would behave
similarly, i.e., both would contribute local magnetic moments and
influence the electric conduction, as well as the
superconductivity.

\begin{figure}
\includegraphics[width=9cm]{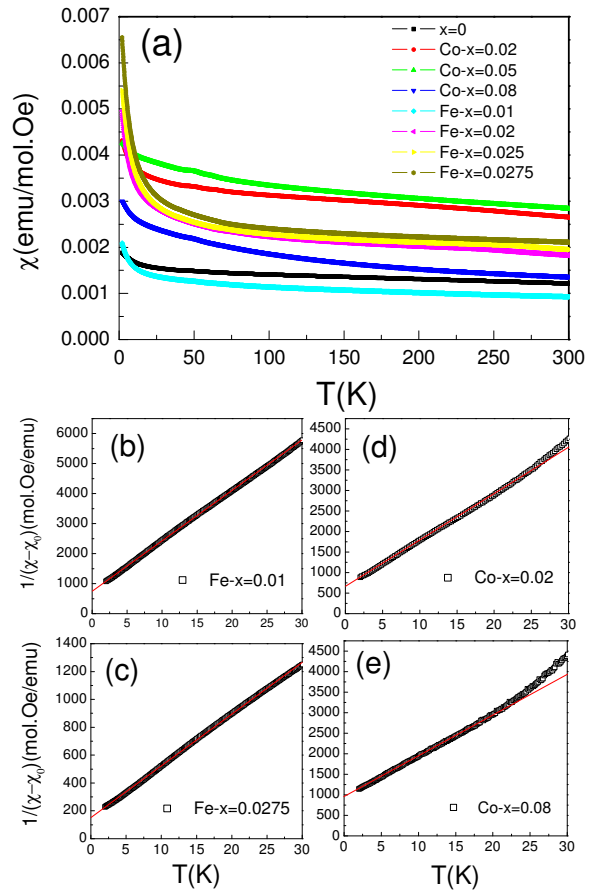}
\caption {(color online) (a) Temperature dependence of DC magnetic
susceptibility for Co and Fe doped samples under 3 T. A low-T
diverging is observed for the Fe-doped samples, indicating an
doping induced local magnetic moments. (b)-(e) The fit to the low
temperature data yielding the magnetic moments (see Table I). }
\label{fig5}
\end{figure}

\begin{figure}
\includegraphics[width=9cm]{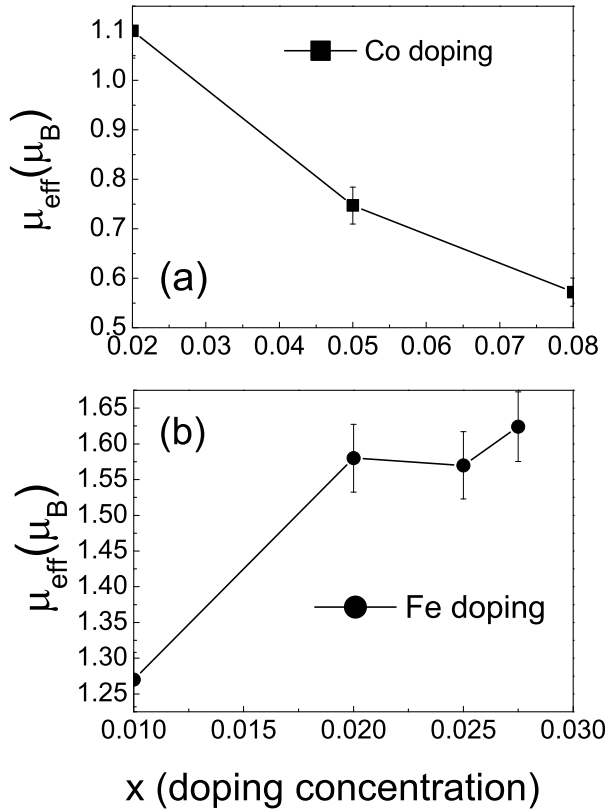}
\caption {(color online) Magnetic moment with Fe and Co doping
calculated by the constant $C$ in the Curie-Weiss law (eq.1). (a)
The doped Fe impurities lead to an enhanced magnetic moment. (b)
The Co doping gives a gradually weakened magnetic moment.}
\label{fig6}
\end{figure}

\begin{table}
\caption{Fitting parameters with Curie-Weiss law for the Co- and
Fe-doped samples. }
\begin{tabular}
{ccccccc}\hline \hline
doping & $C(K\cdot emu/mol \cdot Oe)$ & $T_0(K)$ & $\mu_{eff}(\mu_B$)\\
\hline
Co-0.02 & 0.00883 & 5.596 & 1.1\\
Co-0.05 & 0.01046 & 12.608 & 0.747\\
Co-0.08 & 0.00984 & 9.307 & 0.572\\
Fe-0.01 & 0.00613 & 4.553 & 1.27\\
Fe-0.02 & 0.0188 & 5.097 & 1.58\\
Fe-0.025 & 0.02304 & 4.89 & 1.57\\
Fe-0.0275 & 0.02723 & 4.628 & 1.624\\
 \hline \hline
\end{tabular}

\label{tab.1}
\end{table}

\subsection{Doping induced magnetic moments}
In order to unravel the puzzle concerning the sharp contrast
between the Fe and Co-doped samples, we have done the
magnetization measurements under high magnetic fields. The raw
data of magnetization  measured at 3 T up to room temperature are
shown in Fig.5(a). The temperature dependence of the magnetic
susceptibility look similar, however, it is only for the Fe-doped
samples, that there is a diverging of the magnetic susceptibility
at low temperatures. This diverging of $\chi$ at low temperatures
can be understood as the formation of some strong local magnetic
moments. The magnetization for Co-doped samples reveals an
itinerant moment. To illustrate this point more clearly, we fit
the low temperature magnetization with the Curie-Weiss law,
\begin{equation}
\chi=\chi_0+ C/(T+T_0),
\end{equation}

where $C=\mu_0\mu_{eff}^{2}/3k_B$, $\chi_0$ and $T_0$ are the
fitting parameters. The first term $\chi_0$ arises mainly from the
Pauli paramagnetism of the conduction electrons, the second term
is induced by the local magnetic moments, given by the doped ions.
In order to derive the correct values for $C$ and $\chi_0$, we
adjust $\chi_0$ value to make the $1/(\chi-\chi_0)$ vs. T as a
linear relation in the low temperature limit, the slope gives
$1/C$, and the intercept delivers the value of $T_0$. The data treated
in this way is shown in Fig.5(b)-(e). Here Fig.5(b) and Fig.5(c)
are representing results for the Fe-doped samples with x = 0.01
and x = 0.0275; Fig.5(d) and Fig.5(e) are for the Co-doped ones
for x = 0.02 and x = 0.08. One can see that the low temperature
part is indeed linear. Once $C$ is determined, we can get the
magnetic moment given by the Fe and Co ions $\mu_{eff}$/Co or
$\mu_{eff}$/Fe. It turns out that $\mu_{eff}$/Co =
0.572$\mu_B$ in the Co-doped (x=0.08) sample, 1.62$\mu_B$/(Fe) in
the Fe-doped one (x = 0.0275). Fig.6(a) and Fig.6(b) show the
derived $\mu_{eff}$ for Co- and Fe-doped samples, respectively.
The decrease of the $\mu_{eff}$ in Co-doped samples indicates the
weakening of the magnetic moments compared to the parent sample,
which suggests that the density of states of spin up and spin down contributed by the Co atoms are equal. This is also consistent with
the theoretical results: Co-dopant introduces negligible magnetic
moments. While in Fe-doped samples, an increase of $\mu_{eff}$ is
observed showing the enhancement of magnetic moments by the Fe
impurities. This strongly suggest that the electrons given by the
Fe ions are more polarized, yielding a magnetic moment of about
1.6$\mu_B$/Fe, comparable to the theoretical calculation:
2.05$\mu_B$/Fe.

It is interesting to mention that, although the Ru and Fe are in
the same column in the periodic table, the doped Fe atoms
apparently play a very different role as the Ru does. This is consistent with the common sense that the 3d electrons (here contributed by Fe ions) are more localized leading to the magnetic moments. This is very
different from that in the iron pnictide superconductors in which many
different kind of 3d or 4d transition metals can be doped to the Fe sites
for inducing
superconductivity, showing a wide flexibility.\cite{Mandrus,XuZA,Canfield,HanF,RuDoping} Doping
many transition metals, like Co, Ni, Pd, Ir, Pt and Ru does
not induce very strong magnetic moments, instead the antiferromagnetic order is suppressed. On the other hand, in LaRu$_3$Si$_2$, doping Co does not suppress the superconductivity quickly, although the impurity scattering is strong. This effect manifests that the pairing gap is probably s-wave type, although gap anisotropy exists for the present system.\cite{LiShengPRB} It remains to be explored that whether the Co-doping in LaRu$_3$Si$_2$ can result in a "dome" like doping dependence of superconducting transition temperature, or in other words, can we find an antiferromagnetic (AF) order as the parent phase and superconductivity can be induced by suppressing this AF order.

\begin{figure}
\includegraphics[width=9cm]{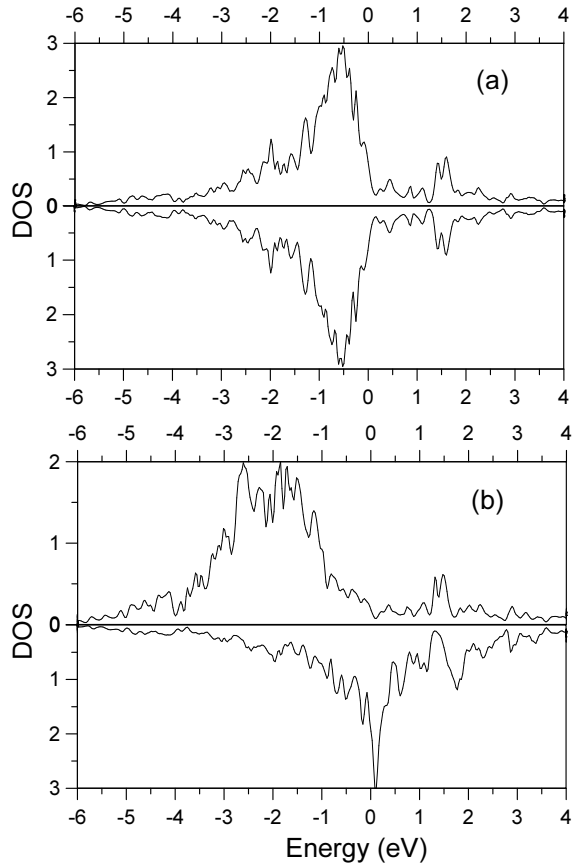}
\caption {(color online) Calculated 3d partial DOS, (a) for Co 3d
orbitals; (b) for Fe 3d orbitals. The positive and negative value
signal the spin up and spin down portion of the DOS.} \label{fig7}
\end{figure}

\subsection{Density-functional theory calculations}

Using WIEN2k package\cite{Blaha}, we studied the electronic
structure based on the generalized gradient
approximation.\cite{Perdew} To consider the low doping
concentration, we perform calculation for a $2\times2\times2$
supercell, and replace one of the 48 Ru atoms in the supercell by
Fe/Co. In Fig.7, we show the Fe/Co 3d partial DOS. It is
interesting to find that the main part of Co 3d is located below
$E_F$. Therefore Co 3d band is close to fully occupied, although
due to the hybridization with Si and Ru, Co 3d has also
distribution above Fermi level ($E_F$). Thus, it is natural to
expect that the spin splitting is very small, and Co becomes
nonmagnetic as shown in Fig.7(a). For Fe, while the spin-up
channel is almost fully occupied like Co, the spin-down is clearly
partially occupied as shown in Fig.7(b). Therefore, there is a big
exchange splitting and the magnetic moment at Fe-site is found to
be 2.05 $\mu_B$, close to our experimental value 1.6 $\mu_B$.
Because of the strong hybridization with Fe 3d electrons, the
neighboring Ru-site has also about 0.1 $\mu_B$ magnetic moment.

\section{Summary}
In summary, contrasting behaviors of the suppression to
superconductivity has been observed in Fe and Co doped
LaRu$_3$Si$_2$. In the case of doping Fe, the superconductivity
can be easily suppressed, while it is much slower in the Co-doped
samples. Measurements and analysis on the DC magnetization suggest
that the Fe-doping induce some strong local magnetic moments,
while Co-doping does not. This is well consistent with our DFT
calculations. In the Fe-doped samples, the impurities act as
strong pair breakers, which is caused by the local magnetic
moment. While the doping of Co atoms brings about equally spin-up and spin-down
electrons which contributes negligible magnetic moment. Therefore
the pair breaking is much weaker in the Co-doped samples.

\begin{acknowledgments}
We appreciate the useful discussions with Jan Zaanen, Zidan Wang,
and Jianxin Li. This work is supported by the NSF of China
(11034011/A0402), the Ministry of Science and Technology of China
(973 projects: 2011CBA00102 and 2012CB821403) and PAPD.
\end{acknowledgments}

$^{\star}$ hhwen@nju.edu.cn

\end{document}